\documentclass{aa}
\usepackage{graphics}
\usepackage{lscape}

\begin{document}

\thesaurus{11(11.19.2; 11.09.5; 11.16.1; 11.09.4; 11.05.1; 13.09.1) }

\title{Spiral and irregular galaxies in the Hubble Deep Field North}
\subtitle{Comparison with early types and implications for the global SFR density}

\author{G.\,Rodighiero\inst{1}
 \and G.L.\,Granato\inst{2}
 \and A.\,Franceschini\inst{1}
 \and G.\,Fasano\inst{2}
 \and L.\,Silva\inst{2} }

\institute{ Dipartimento di Astronomia di Padova, Vicolo dell'Osservatorio, 5, I-35122 Padova, ITALY
 \and Osservatorio Astronomico di Padova, Vicolo dell'Osservatorio, 5, I-35122 Padova, ITALY}
\offprints{G. Rodighiero}
\mail{rodighiero@pd.astro.it}
\titlerunning{Spiral and Irregular Galaxies in the HDF North}
\date{Received  / Accepted}
\maketitle

\begin{abstract}

\bigskip

We analyze a morphologically-selected complete sample of 52 late-type 
(spiral and irregular)
galaxies in the Hubble Deep Field North with total K-magnitudes
brighter than K=20.47 and typical redshifts $z\sim 0.5$ to 1.4. 
This sample exploits in particular the
ultimate imaging quality achieved by HST in this field, allowing
us to clearly disentangle the early- from late-type galaxy morphologies,
based on accurate profiles of the surface brightness distributions.
Our purpose was to investigate systematic differences between the
two classes, as for colours, redshift distributions and ages
of the dominant stellar populations.
Our analysis makes also use of an exhaustive set of modellistic spectra
accounting for a variety of physical and geometrical situations
for the stellar populations, the dusty Interstellar Medium (ISM), and 
relative assemblies.
The high photometric quality and wide spectral coverage
allow to estimate accurate photometric
redshifts for 16 objects lacking a spectroscopic measurement, and allow
a careful evaluation of all systematics  of the selection
[e.g. that due to the surface-brightness limit]. 
This sample appears to miss significantly
galaxies above $z=1.4$ (in a similar way as an early-type galaxy sample 
previously studied by us), a fact which may be explained as a global
decline of the underlying mass function for galaxies at these high redshifts.
Differences between early- and late-types are apparent
-- particularly in the colour distributions and the evolutionary 
star-formation (SF) rates per unit volume --, although the complication 
in spectro-photometric modelling introduced by dust-extinction 
in the gas-rich systems prevents us to reach conclusive results
on the single sources (only future long-wavelength 
IR observations will allow to break the age/extinction degeneracy).
However, we find that an integrated quantity like the 
comoving star-formation rate density as a function of redshift  
$\Psi(z)$ is much less affected by these uncertainties: by combining this with
the previously studied early-type galaxy sample, we find a shallower 
dependence of $\Psi(z)$ on $z$ between $z=0.2$ and $z=1.5$ than found by 
Lilly et al. (\cite{lilly}). Our present results, based on a careful modelling 
of the UV-optical-NIR SED of a complete galaxy sample --
exploiting the observed time-dependent baryonic mass function in stars as a 
constraint and attempting a first-order correction for dust extinction 
-- support a revision of the Lilly-Madau plot at low-redshifts 
for both UV- and K-band selected samples, as suggested
by independent authors (Cowie et al. \cite{cowie}).


\end{abstract}

\bigskip 

\section{INTRODUCTION}

Cosmogonic models, in particular those based on the hierarchical
clustering of cold dark matter halos, now including detailed
physical descriptions of gas cooling, star-formation and feed-back
processes in the baryonic component,
make specific predictions about the evolutionary history of galaxy
populations as a function of their morphology.

Basically, in the hierarchical scheme, forming galaxies acquire
angular momentum from tidal interactions with the surrounding
structure and then dissipate and collapse preferentially along the rotation 
vector and tend to form flattened rotational-supported structures 
(disk galaxies). There are indeed indications of a
substantial population of large structures of this kind up
to the highest redshifts from absorption-line studies in the
distant quasar spectra (e.g. Wolfe A.M. \cite{wolfe}).
The pressure-supported stellar bulges dominating E/S0 galaxies, in this scheme,
would originate from the violent relaxation and dynamical evolution 
following strong interactions and mergers  of
primordial disk galaxies. At the zero-th order, these models 
predict that spheroidal galaxies are assembled somewhat later than spirals,
although their stellar populations might not differ much if the
merger occurs among gas-poor systems. One such extreme case
has been discussed by Kauffmann \& Charlot (\cite{kauffmannb}) based on the $\Omega=1$ 
standard CDM
cosmology, predicting a substantial dearth of spheroidal galaxies
already by z=1, most of them being formed at lower z.

More recently it has been pointed out that, within this scheme,
the morphological appearence of a galaxy
may repeatedly change with cosmic time not only from a late- to an 
early-type following a merging event, but also from a spheroidal
to a disk configuration following the acquisition of new infalling
gas from the environment (e.g. Ellis R.S. \cite{ellis}). This may reflect in
long-wavelength (V to K) colour distributions un-distinguishable
between early- and late-types, while shorter-wavelength (U to V) 
colours would be dominated by the on-going SF in disks. 

An opposite pattern is contemplated by the "traditional" models of
galaxy formation, assuming that massive galaxies, in particular 
elliptical and S0's, originated first at high redshifts as single entities
from rapid homologous collapse of primordial gas. Gas-rich systems, in
this view, form instead more quiescently from progressive inflow of gas into
the dark matter halos during most of the Hubble time. 
This formation scenario then predicts a marked
differentiation in colours and ages for the stellar populations of
the two classes of galaxies, late-type galaxies containing much younger
stellar populations on average than the early-types. Also a substantial
population of massive spheroids would be expected to be visible at
$z\sim 2$ in this case.

The ultra-deep integrations at various wavelengths performed 
by the Hubble Space Telescope in the Hubble Deep Field (Williams et al. 
\cite{williams}; 
we consider here only the survey in the North area) offer an extremely valuable 
dataset to study morphological properties of high-redshift galaxies.
Furthermore, the very accurate photometry achievable in such deep images
allows accurate estimates of the photometric redshifts for vast numbers of 
faint galaxies in the field.

We have recently exploited these data to study the colours, masses, age
distributions, and the star-formation history of a sample of elliptical-S0 galaxies
(Franceschini et al.  \cite{franceschini}, FA98 hereafter). 
The basic result was to find colours indicative
of wide ranges of ages for the stellar populations and a remarkable
absence of objects at $z>1.3$, both facts telling against the predictions
of the "traditional" monolithic formation scenario.

As a natural complement, we present in this paper an analysis of late-type
and irregular galaxies in the HDF. Similarly to what we did there, our
primary selection is in the K-band, obtained from a deep KPNO image,
to minimize the biases in the sample due to the 
effects of K- and evolutionary corrections.
The completion of our previous analysis of E/S0 to account for the 
complementary set of late-type systems is also needed for a global 
evaluation of the star-formation history as a function of redshift.
The advantage of our approach over previous attempts (Lilly et al. 
\cite{lillyb}, 
Cowie et al. \cite{cowie}) is in our careful treatment of dust extinction
from a detailed fitting of the UV-optical-NIR spectral energy distributions
(SED). In addition, the detailed knlowledge of the 
near-IR (NIR) spectrum for sources
at the relevant redshifts is informative on the baryonic mass function in
stars, which provides an essential constraint on the cumulative 
star-formation rate as a function of time.

In Section 2 we discuss the selection
scheme and photometric corrections  used to construct a complete K-band flux limited sample of late--type galaxies.
In Section 3 we describe the population synthesis code 
that we used to model the optical-NIR 
SEDs of our sample objects, taking into full account 
the effects of a dusty interstellar medium in the galaxy spectra.
Our main results are then reported in Section 4, where we 
perform detailed analyses of the space distributions, 
colours and ages of the stellar populations of field spiral and
irregular galaxies, compared with ellipticals and S0.
We discuss the difficulties inherent in the spectral
modelling of gas-rich systems affected by dust extinction.
We finally attempt to construct the global star formation histories of
field galaxies (E/S0+spiral/irregulars).
In Section 5 we summarize our main conclusions.

We anticipate that the results of the  present analysis, 
based on a survey over a very small sky field, are to be
considered as only tentative, untill larger areas will be surveyed
to similarly deep limits.

We adopt $H_0=50~ Km~ s^{-1}~ Mpc^{-1}$ throughout the paper. 
For consistency with FA98 the analysis is made assuming
$q_0=0.5$, and zero cosmological constant $\Lambda$.

\section{SAMPLE SELECTION AND PHOTOMETRY}

The Hubble Deep Field North has been observed in 4 broad bands (F300W, 
F450W, F606W, F814W) for a total of 150 HST orbits by Williams et al. (1996),
and constitutes the deepest ever exposure on a small sky area.
Accurate photometric data in the four bands have been published
for thousands of faint galaxies by the authors.

Dickinson et al. (\cite{dickinson}) observed the HDF--North in the near-IR
with the IRIM camera on the KPNO 4 m telescope.
The camera employs a 256 $\times$  256 NICMOS-3 array with 
0".16 pixel$^{-1}$, but the released images were 
geometrically transformed and rebinned into a 1024 $\times$
1024 format. IRIM exposures have been secured in the $J$, 
$H$ and $K$ filters, for a total of 12, 11.5 and 23 hours,
respectively. Formal 5 $\sigma$ limiting magnitudes for the
HDF/IRIM images, computed from the measured sky noise
within a 2" diameter circular aperture, are 23.45 mag at $J$,
22.29 mag at $H$, and 21.92 mag at $K$, whereas the image
quality is $\sim$ 1".0 FWHM.

Our sample of galaxies has been extracted from the HDF/IRIM $K$-band
image through a preliminary selection based on the automatic
photometry provided by SExtractor (Bertin $\&$ Arnouts \cite{bertin}). It
is flux limited in the $K$ band and it excludes early--type
galaxies, i.e. objects whose surface brightness distribution is
dominated by a de Vaucouleurs profile. 

To determine the limit of completeness in the $K$ band (hereafter
$K_L$) for inclusion in our sample, we followed the same empirical
procedure described in FA98. Briefly, a large
number of toy galaxies with exponential profiles (produced with the
IRAF-MKOBJECTS tool) have been used to check the performances of
Sextractor in estimating the K-band magnitudes ($K_{SEx}$) of
late-type galaxies. This allowed us to determine the magnitude
$K_{lim}$ below which the scatter $\sigma_{K_{SEx}}$ of the measured
magnitudes turns out to be lower than some given value $\sigma_{max}$
(the vertical line in Figure 1a corresponds to $K_{lim}=21$ and
$\sigma_{max}=0.18$). Moreover, using only galaxies with $<K_{SEx}>\le
K_{lim}$, we have derived the following empirical relation between the bias
$\Delta K=<K_{SEx}>-K_{true}$ and the effective surface brightness
$<\mu_e^K(SEx)>$ (see Figure 1b):
\begin{eqnarray}
 & \log \Delta K=-1.059+0.188<\mu_*^K>+0.061<\mu_*^K>^2 
  \nonumber\\
   &+0.006<\mu_*^K>^3,
\end{eqnarray}
where $<\mu_*^K>=<\mu_e^K(SEx)>-21$. This relation provides the true 
total $K$ magnitudes from the SExtractor measured flux. Then simulations
have shown that the K-band image has a surface brightness limit
of $<\mu_e^K(SEx)>=23.$

\begin{figure}
\begin{center}
 \resizebox{9cm}{!}{\includegraphics{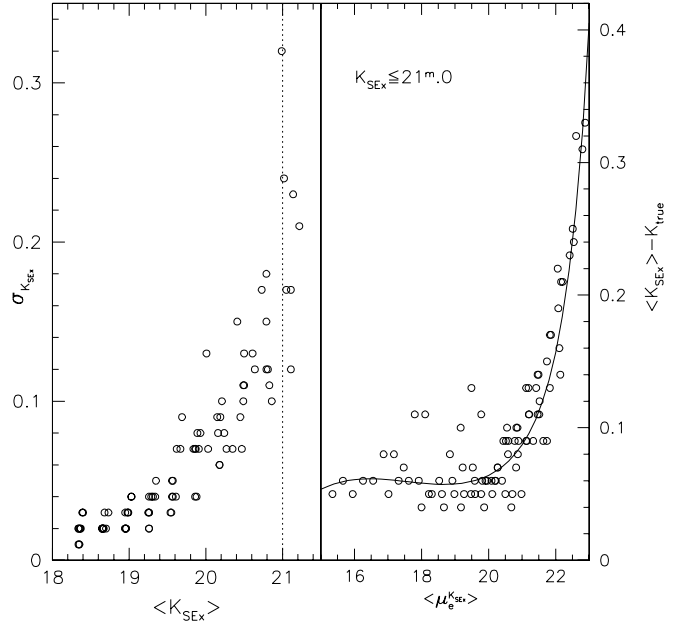}}
  \caption{Left panel: standard deviation of 
 the SExtractor magnitude estimates as a function of
 the average magnitude for galaxies in the simulated images: the standard deviation
 is very small for $<K_{SEx}>\le 21^m.0$. Right panel: difference between true flux and the
SExtractor flux as a function of the average surface brightness for the subsample
of simulated galaxies with  $<K_{SEx}>\le 21^m.0$.}
 \label{figfitk}
\end{center}
\end{figure}

By analogy with FA98, we have first produced a catalog of
morphologically selected late-type objects with $K_{SEx}\le
K_{lim}=21.0$; then we have used effective radius estimates from high
resolution (HST) optical imaging to derive the effective surface
brightness of galaxies ($<\mu_e^K>$); finally, we have applied to the
$K_{SEx}$ magnitudes the statistical corrections given by equation [1]
and we have included in the final sample only galaxies with corrected
magnitudes less than or equal to $K_L=K_{lim}- \Delta K_{max}-
\sigma_{max}=20.47$ mag (we assume $\Delta K_{max}\sim 0.35$ from Figure 1b).

\begin{figure*}
\begin{center}
 \resizebox{16cm}{!}{\includegraphics{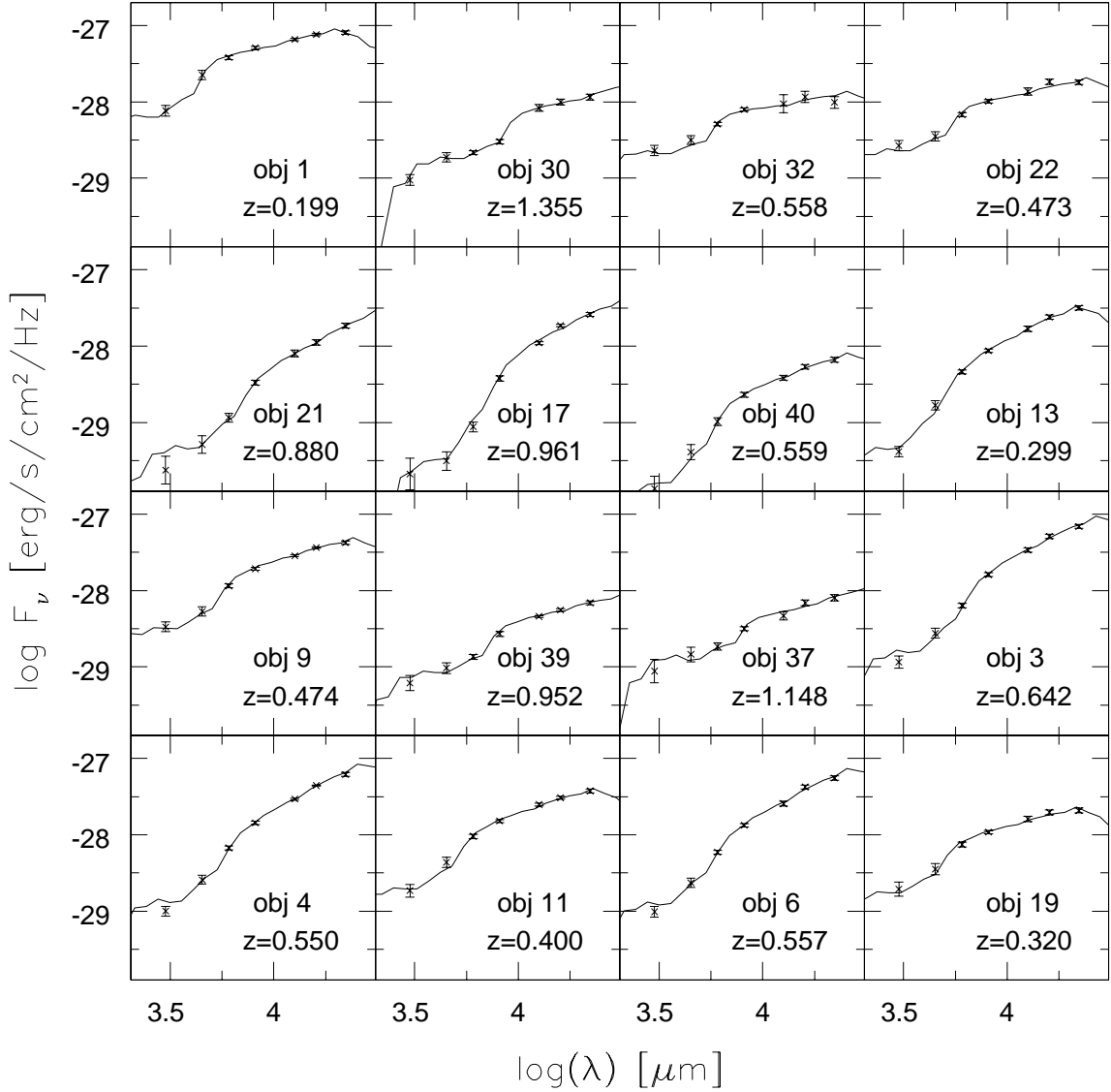}}
\caption{Observed broadband spectra for sixteen galaxies in our sample, 
fitted with the synthetic models described in Section 3.}
\label{fit_grid}
\end{center}
\end{figure*}

A total of 176 objects with $K_{SEx}\le 21$ mag were detected by 
SExtractor in the IRIM $K$-band image. After carefull inspection
of the high resolution HST images, we rejected all elliptical and 
S0 galaxies (including the 35 early-type objects identified in 
FA98) and the stars.
A few objects were also rejected from the sample due to their 
position in the frame (at the edges of the image the noise is
higher and the magnitude estimate is likely to be uncertain).
We then produced a first preliminary, incomplete sample of late--type 
galaxies.

The effective radii $r_e$ were estimated running SExtractor
on the WFPC2 $V_{606}$ frame with the  parameter
$FLUX\_RADIUS$, providing the radius containing half of the 
total emitted flux. The surface
brightness $<\mu_e^K(SEx)>$ was evaluated for each galaxy and the 
statistical corrections $\Delta K$ were computed using the eq (1). 

The final complete sample of late-type galaxies with $K_{corr}<20.47$ 
consists of 52 objects over the HDF area of 5.7 square arcmin. 
For 36 objects we have the spectroscopic redshift (Cohen et al.  
\cite{cohen}, 
Cowie et al. \cite{cowie}, Fernandez-Soto et al. \cite{soto}),
while for the remaining 16 we measured it from our photometric analysis
as described below.

A procedure analogous to that outlined for the K band was used to
derive, for each object of the selected sample, the corrected 
magnitudes in the J and H bands. 

The optical magnitudes in the F300W, F450W, F606W and F814W bands (U,
B, V and I in Table 1, respectively) have been computed again with
SExtractor on the high resolution WFPC2 images (no
corrections being applied in this case). Magnitudes are in the AB
system, defined by the relation (Oke $\&$ Gunn \cite{oke}):
\begin{equation}
 AB=-2.5\log F_{\nu}-48.60
\end{equation}
where $F_{\nu}$ is the flux in $erg~cm^{-2}~Hz^{-1}$, the constant
beeing choosen so that $AB=V$ for an object with flat spectrum.

Some data on the sample are listed in Table 1. Column 1: our 
identification; 
column 2-4: coordinates $\alpha$ and $\delta$ (at J2000). 
To these must be added 12 hours 36 minutes (RA) and 62 degrees (Dec);
column 5: $r_e$ is the effective radius, in arcsec, derived from HST images; 
column 6-9: optical U, B, V, I magnitudes in the AB system (see above);\
column 10-12: near-infrared J, H, K corrected magnitudes in the standard system; 
column 13: redshift of each object. Values in brackets are
 photometric redshifts, while the other are all spectroscopic.

\section{MODELLING GALAXY SEDs IN THE PRESENCE OF A DUSTY ISM}

The optical-NIR SEDs of our sample objects have been modelled using the 
population synthesis code GRASIL (Silva et al.\ \cite{silva}), taking into full account
the effects (optical extinction and thermal reprocessing) 
of a dusty interstellar medium in galaxy spectra. 
We defer the reader to that
paper for a through description of this model and for precise definitions of the
parameter, while for convenience we summarize the main features below.

\subsection{The GRASIL code}

The code provides a self-consistent description of the formation and
evolution of a galactic system in its various stellar and ISM components, including its 
secular evolution during the Hubble time and episods of enhanced star-formation
possibly following interactions and mergers.

As a preliminary step the code allows to solve the equations ruling the chemical 
evolution, providing the star formation and metallicity histories
SFR(t) and Z(t) as a function of time. The computations presented here were
performed adopting one--zone (no spatial dependence) open models
including the infall of primordial gas, according to the standard equations of
galactic chemical evolution. As usual, the star formation rate is determined by the
amount of gas in the system according to a Schmidt-type law
\footnote{The code can be downloaded from {\it
http://grana.pd.astro.it} }
$$SFR(t) = \nu \, M_g(t)^k .$$
We have generated 3 different $SFR(t)$, in order to provide a wide
range of spectral evolution patterns.
The peak occurs at about 1, 2 and 3 Gyr (hereafter model 
(a), (b) and  (c) respectively), getting broader from (a) to (c). 
As a result, half of
the final stellar mass $M$ (i.e.\ at 13 Gyr) has been assembled at galactic times
of 2, 3.7 and 4.7 Gyr in the three cases respectively. A standard Salpeter
IMF between 0.1 and 100 $M_{\odot}$ is assumed.

As described by Silva et al. (1998), GRASIL calculates self-consistently the
absorption of starlight by dust, the heating and thermal emission of dust grains, 
for an assumed geometrical distribution of the
stars and dust, and a specific grain model.

In the GRASIL model several parameters affect the overall
modifications imprinted by dust on the SED. However, if we confine
ourselves to the attenuation of stellar radiation in the optical/UV/NIR bands, we can
obtain most of the possible spectral behaviours by adjusting only two quantities:
the $escape~timescale~t_o$ (see Eq [8] in Silva et al. 1998 for a precise 
definition) of newly formed stars from parent molecular 
clouds (MCs) and the total mass of dust. 
Indeed $t_o$ controls the fraction of light from very young stellar generations
hidden inside MCs and converted to IR photons, since the MCs optical
thickness is very high below $\sim 1 \mu$m (cfr.\ Silva et al.\ 1998).
On the other hand, the effects of the diffuse (cirrus) dust depend on
several quantities: the radial and vertical scale lengths for stars and
dust distributions $R_d$ and $z_d$, the residual gas in the galaxy
$M_g$, the dust to gas ratio $\delta$ and the fraction of gas which is
in the MCs component $f_{mc}$. However we found that most, if not all,
the possible attenuation laws of the diffuse dust, arising from
different choices of these quantities, can be closely mimicked by simply
adjusting the amount of gas, while fixing the other quantities to the
`typical' values: $R_d = 3.5 (M/10^{11}M_\odot)^{1/3}$ Kpc,
$z_d=0.1 \, R_d$, $\delta=0.01$ and $f_{mc}=0.5$. Obviously,
while different choices of  $M_g$, $R_d$, $z_d$, $\delta$ and
$f_{mc}$ can yield similar attenuation laws on the optical spectrum, the
spectral shapes of the corresponding IR continuum re-radiation 
can be rather different.

\begin{figure}
\begin{center}
 \resizebox{9cm}{!}{\includegraphics{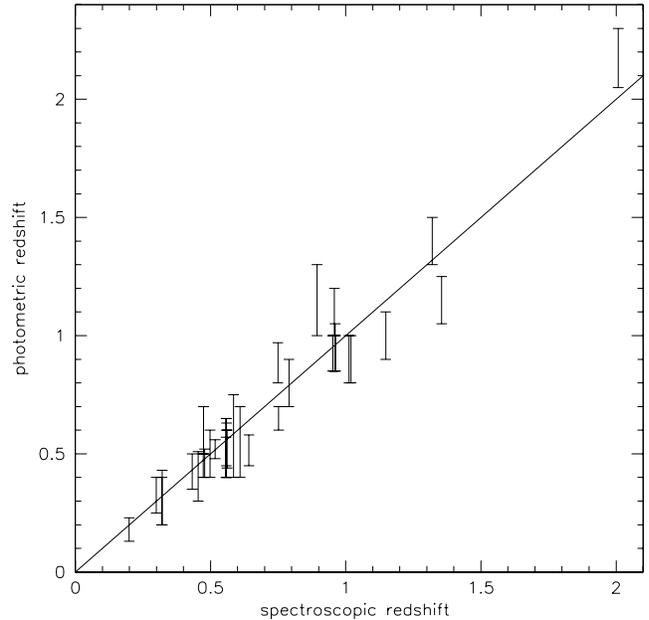}}
\end{center}
\caption{Comparison of photometric redshifts, based on seven-band spectral
data, with spectroscopic redshifts. Error bars
refer to interval solutions with more than 90$\%$ confidence level,
derived from $\chi^2$ fitting using models [a] and [c].}
\label{zspphot}
\end{figure}

Strictly speaking the residual gas $M_g$ is not a parameter, being
instead the outcome of the chemical evolution code,
through the Schmidt law. However we use the
trick of forcing $M_g$ to different values, in order to describe with a
monoparametric sequence the effects of a global attenuation on
the SED. Besides this, a larger `freedom' on $M_g$ takes into account
that the Schmidt law should not be taken too literally, as a strict
relationship between the total gas content and the $SFR$ in the system.
The law may only provide an order of magnitude description, in
particular for the secular evolution of the SFR, the so-called "inactive
phase" of galaxy evolution bringing essentially to the formation of
spiral disks. Several other physical parameters influence the rate of
star-formation with respect to the simple available amount of residual
gas, in particular the gas pressure and temperature, which may
drastically change as a consequence of a violent dynamical event, like
an interaction or a merger, followed by gas compression and efficient
cooling.
Overall, we use the criterion of considering acceptable
values from 0.2 to 5 times the `true' $M_g$ given by the chemical
evolution code.

\subsection{An extensive grid of model template spectra}

The code allowed us to build a very large set of model spectra describing
all possible age and mass distributions for the stellar populations, for
the dusty ISM, and relative assemblies.

For each of the 3 histories SFR(t) we have generated two grids of models: one
with $t_o=5$ Myr and another with $t_o=30$ Myr. Silva et al.\ (1998) found that
the former value is typical for normal spirals while the latter is more suited for
starbursting systems. Each of these grids consists of 1400 models computed with
ages $t_G$ ranging from 0.2 to 10 Gyr in steps of 0.2 Gyr and $M_g$ from 0 to 1
(in units of the final mass of stars) in 28 logarithmic steps. 

In total we have
therefore $1400\times 2 \times 3 = 8400$ model spectra with different age, gas content,
MCs escape timescale, and $SFR(t)$ which we compared with the observed sample SED,
allowing for the obvious scaling in luminosity. 

In addition we considered one further grid of spectra to see how our observed SEDs compare with those expected for
spheroidal systems: for these we used the $SFR(t)$ (c) model, but
truncated at 3 Gyr to simulate the onset of a galactic wind. The adopted geometry in
this case was a modified King profile (Eq.\ [3] in Silva et al.\ 1998) with $r_c =
0.15 (M/10^{11}M_\odot)^{1/3}$ Kpc.

An example of the resulting fits to the observed broadband spectra of sixsteen
galaxies in our sample is reported in Figure\ \ref{fit_grid}.
The analysis of the 52 fitted SEDs reveals the presence of 
two dominant different kinds of spectral behaviours:
(a) objects which are red and show a strong convergence
in the UV region, and (b) blue spectra that are flatter at all
wavelengths, dominated by young stellar populations.

\begin{figure}
\begin{center}
 \resizebox{9cm}{!}{\includegraphics{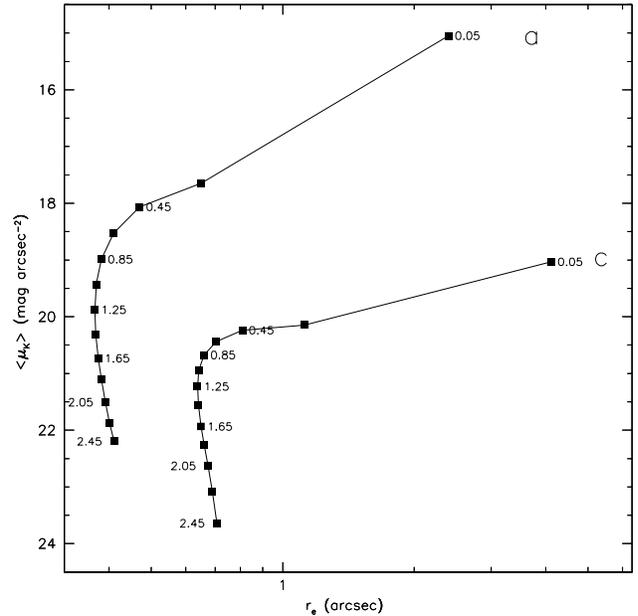}}
\end{center}
\caption{Scaling of the observed average surface brightness for two faint 
galaxies
in our sample as a function of redshift, according to the evolution models (a) 
for galaxy \# 16 and (c) for \# 50 respectively. 
Note that galaxy \# 50 is the one with the lowest observed value of the surface brightness.
The deep K-band image used in the primary selection has a $3\sigma$ limiting
brightness of $23\ K\ magnitudes/arcsec^2$.}
\label{figsurf}
\end{figure}

\begin{figure}
\begin{center}
 \resizebox{\hsize}{!}{\includegraphics{9985.f5}}
\end{center}
\caption{Redshift distribution for spiral and irregular galaxies brighter
than $K=20.47$ in the HDFN.  The continuous line is the predicted distribution
based on canonical local luminosity functions for late-type and irregular
galaxies and the spectral evolution model (c) mentioned in Sect. 3.1. }
\label{figred}
\end{figure}

\begin{figure*}
\begin{center}
 \resizebox{12cm}{!}{\includegraphics{9985.f6}}
\caption{Examples of different fits to the observed broad-band spectra of object
number 1, fitted with the population synthesis  models described in Sect. 3. The
figure plots $\chi^2$ contours for the amount of dust-rich gas versus the age
(top panels) and the corresponding best-fit spectra for two different 
histories of SF (lower panels). The solution on the right refers to the model [a]
in Fig. 2 and a value for the escape timescale $t_o$ of 5 Myr (see Sect. 3.1 for details
about these parameters). On the left: solution for model [c] and $t_o= 30$ Myr.
The contours correspond to a $\chi^2$ increment of 1,5,10,15 and 20 with respect
to the best fit in the grid. Therefore the second innermost contour 
correspond to the 90\% confidence interval.}
\label{figchi}
\end{center}
\end{figure*}

\begin{figure*}
 \resizebox{8cm}{!}{\includegraphics{9985.f7}}
 \resizebox{8cm}{!}{\includegraphics{9985.f8}} 
 \hfill 
 \resizebox{8cm}{!}{\includegraphics{9985.f9}} 
\caption{For three representative objects in our sample ($\#$ 1, 30 and 36) 
located at different redshift (z = 0.199, 1.355 and 0.559 respectively) we
plot various solutions for the rate of on-going SF against the average V-band
effective extinction $A_V$. We report the best-fit solutions for every different
history of SF considered (as described in Sect. 3.2) and two more
corresponding models: the youngest and the oldest one with $\chi^2=
\chi^2_{best}+5$. The labels near the points are the  
$\chi^2$ values for each solution. 
This figure illustrates the fact that the observed SEDs can be fitted with
models differing in the current rate of star formation by factors up to 5-10.}
\label{figavsf}
\end{figure*}

\section{RESULTS}

\subsection{The space-time distribution of K-band selected late-type galaxies}

The grid of model spectra has been used to estimate redshifts from
spectral fits to the 7-band photometric data for the 16 galaxies lacking
a spectroscopic measurement. As typical in cases 
in which such a wide spectral coverage (0.3 to 2.2 $\mu m$) and accurate
photometry are available, the relative errors in $z$ turn out to be quite 
small, of the order of $\sim 10\%$.
These broad-band spectral fits allow quite robust estimates of redshifts
also for dusty objects, 
mostly exploiting a well-characterized feature of the optical spectra, 
the Balmer discontinuity, which is weakly affected by dust extinction
(cfr discussion in section 3.2.1 in FA98).

\begin{figure}
\begin{center}
 \resizebox{9cm}{!}{\includegraphics{9985.f10}}
\end{center}
\caption{Example of two different fits to the observed broad-band spectrum for
object number 30 in our identification. We report the spectral galaxy emission 
of two dusty environments (Sect. 3.1): dashed line = diffuse ISM (cirrus),
dot-dashed line = molecular clouds (MCs). The solid line corresponds to the
total integrated spectrum of the galaxy. Solution 1 (top panel) and 2 (bottom
panel) refer to that plotted in Fig. 7 for object 30, corresponding to a value 
of $A_V \sim 1.9$, with SFR $\sim 180 M_{\odot}/yr$, $\chi^2=6.77$ (for sol1), and SFR $\sim 420 M_{\odot}/yr$, $\chi^2=7.44$ (for sol2). }
\label{30fig1}
\end{figure}

To check the consistency of our method, we compare in Figure\ \ref{zspphot}
our photometric redshift predictions with
the corresponding spectroscopic measures. The vertical error bars
refer to different solutions at better than 90$\%$ confidence,
derived from a $\chi^2$ fitting procedure using models (a) and (c).
Fig.\  \ref{zspphot} shows overall good agreement within our modellistic
uncertainties of the fits.
Added to the 36 spectroscopic redshifts, this procedure enabled us to get quite a
reliable redshift distribution for our sample objects.

The distribution in redshift of a source population from a complete
flux-limited catalogue provides a powerful
constraint on its evolutionary history and formation epoch.
This obviously assumes that we control with reasonable confidence all
possible selection effects, in particular those due to the surface 
brightness limit, the cosmological dimming and K-corrections to the fluxes.
If some morphological criteria are at play, one needs also to understand 
how morphological appearence may evolve with redshift.
In our case the control of the selection effects is made easier by our
primary selection in the K-band, which implies minimal K- and evolutionary
corrections as a function of redshift.

The availability of accurate measurements of the effective radii $r_e$
allows to control the effects of the limiting surface brightness
observable in the field in K, which we evaluated from the simulations described in 
Sect. 2 to be $<\mu_K>\simeq 23\ magnitudes/arcsec^2$.

Figure\ \ref{figsurf} shows the evolution of the observed surface brightness for two
objects in our sample (including the galaxy with the faintest surface brightness) 
as a function of redshift, taking into account the cosmic dimming
and using a variety of spectral evolution patterns corresponding to the models
described in Sect. 3.
It is clear from the figure that the cutoff in surface brightness in the IRIM
K band image has no impact in our selection process above our adopted limit
in total magnitude of $K<20.47$, and the whole redshift space up to at least
$z=2.5$ is clearly accessible in principle.

Figure\ \ref{figred} reports the histogram of the observed redshifts for our
complete $K<20.47$ sample (including the 16 photometric estimates).  
The distribution shows pronounced peaks at $z\sim 0.5$ and $z\sim 1$, clearly
indicative of strong inhomogeneities in the source distribution due to
spatial clustering in the relatively small volume sampled by the HDFN.
The uncertainty due to clustering in the limited volume 
has to be kept in mind when drawing any conclusions from our analysis, which
require confirmation from surveys on more extended areas.

A second relevant feature is apparent in Fig.\ \ref{figred}:
a cutoff at $z\sim 1.4$, with only 2 objects out of 52 found above this limit. 

We compared this distribution with a model
prediction based on the local luminosity function of galaxies in the
K-band (Gardner et al. \cite{gardner}) 
complemented with information on the contributions
of various morphological classes from optical data (see Franceschini et al. 1998
for more details). The luminosity function is then evolved according to spectral 
model (c), which provides a conservative estimate of the number of $z>1$ galaxies
(it has the minimal evolution rates among the three models considered). 
The other assumption we made is that the luminosity function changes
as an effect of the evolution of the $M/L$ ratio ($M$ changes because
more mass is turned into stars with time, $L$ follows the evolution of the
stellar populations). The onset
of star-formation is assumed to happen at $z=4$. 

As shown in the figure, these assumptions would imply an expected number
of late-type galaxies at $z>1.5$ significantly in excess of the observations
(9 expected versus only 1 observed).
This result parallels a similar finding by FA98 and Rodighiero, Franceschini and
Fasano (2000) for the early-type population,
showing a demise of objects at redshifts larger than $z\sim 1.3$.
One of the possible interpretations of this effect given by FA98, i.e. that the morphological
selection could miss galaxies with shapes deviating from the De Vaucouleurs
profile in case of merging activity at these redshifts, is no longer
acceptable: essentially there are no bright ($K<20.5$) galaxies altogether 
at $z>1.4$ in the HDFN area.

\subsection{Evaluating galactic ages and extinction properties}

As anticipated, if the photometric measurement of redshift from
broad-band spectral fits is weakly affected by dust extinction,
the estimates of most other physical parameters of gas-rich systems
suffer quite more by the uncertain amount of dust and from the 
degeneracy between ages of stellar populations and extinction. 
To check this we have used our large grid of model spectra to study 
the degeneracy of the spectral-fitting solutions of
our sample galaxies.

Figures \ref{figchi} plots $\chi^2$ contours for the amount of
dust-rich gas versus the age for two representative galaxies, as well as the
best-fit spectra for two different histories of SF (model [a] and [c]).

It is immediately apparent that, even within sets of models based 
on the same evolutionary SFR(t), a fairly substantial
 degeneracy exists between the age and amount of dust. 
Furthermore,
rather different star formation histories can lead to equally good fits,
as detailed in Fig.\ \ref{figchi}.

\begin{figure*}
 \begin{center}
 \resizebox{13cm}{!}{\includegraphics{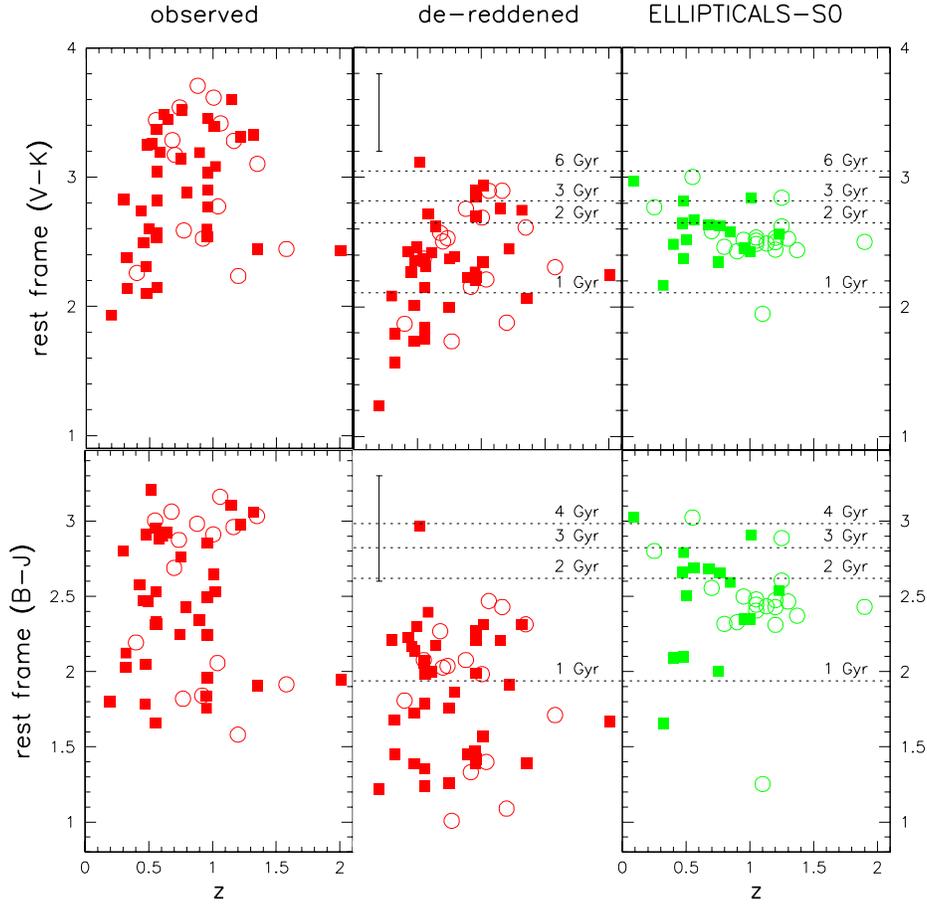}}
\caption{Rest frame (V-K) (upper panels) and (B-J) (lower panels) colours
of late--type field galaxies,
compared with predicted values for single stellar populations with solar
metallicity. The ages for the latter are indicated as well the mean colours
of local galaxies.
The left two panels refer to the observed colours,
on the center we present the corresponding
"de-reddened" colours based on our best-fit SED solutions.  
On the right we report the rest-frame colours of
ellipticals and S0 galaxies from FA98.
Filled squares refer to objects with spectroscopic
redshift, open circles to those with photometric redshift.
The error bars shown in the central panels correspond to the
uncertainty in the dereddening at 90\% confidence. }
\label{color}
\end{center}
\end{figure*}

Figure \ref{figavsf} summarizes some results of our best-fitting procedures
for three representative objects in our sample. For each object, 
it reports various solutions for the rate of on-going SF and the average
V-band extinction $A_V$, including the corresponding values of the
$\chi^2$. This figure illustrates the fact that the observed SEDs can be fitted 
with models differing in the current rate of star-formation by factors up to 
5--10: 
a large amount of SF activity can be easily hidden at wavelengths below a
 few $\mu$m. 

Figure\ \ref{30fig1} details the results of two different
fits to the observed broadband spectrum for object 30, clearly
illustrating the degeneracy existing between SF and extinction. The two
SEDs correspond to two solutions reported in Fig.\ \ref{figavsf}, 
with values of the SFR differing by a factor $\sim 2.5$.
The top panel refers to the solution 1 with $\chi^2=6.77$ and 
SFR$\sim 200M_{\odot}/yr$. The lower panel refers to solution 2
 with $\chi^2=7.44$, SFR$\sim 400M_{\odot}/yr$. 
It is clear that if the analysis is confined to optical/NIR wavelenghts, 
it cannot clearly discriminate
between the two solutions, whose differences are apparent only including
the far infrared spectrum, where dust re-emission would be detectable.
Only observations of the IR spectral energy distribution, say
between a few tenths up to a few hundreds $\mu$m, where actively star
forming galaxies emit most of the energy, would allow to break the 
present degeneracy in the solutions.

\subsection{Colours, sizes and average surface brightness of late-type galaxies 
at high redshifts}

As a first assessment of the age and extinction distributions,
we report in Figure\ \ref{color} the rest frame (V-K) and (B-J)
colours as a function of redshift. As in FA98, the rest frame
(B-J) colours are computed by interpolating the observed galaxy
spectra using the best-fit models listed in Table 1 (see Sect. 4.4),
while the (V-K) colours require a slight extrapolation to longer
wavelenghts. The left two panels refer to the observed spectra,
which include the effects of reddening, while the colour distributions
reported in the central panels correspond to  
"de-reddened" spectra (i.e. taking out the effect of extinction and
showing the underlying colour distribution). As we see, 
extinction plays a significant role: the estimated 
absorption-subtracted colours appear on average bluer by one magnitude.
De-reddened colours are compared with the predictions
of single stellar populations with solar metal abundances
 (dashed horizontal lines).
The vertical error bars in the central panels are the mean 
uncertainties related
to our de-reddening procedure based on our grid of models and our
adopted 90\% confidence level.

A comparison with the early--type galaxy sample studied by FA98
indicates that our late--type field galaxies present redder colours on 
average, because of extinction. This evidence is stronger in the (V-K)
distribution, where a remarkable excess of red late--types is
apparent at $z>0.5$ and $V-K>3$.  This illustrates that selecting by colours
is far more sensitive to extinction effects than to intrinsic
differences among the stellar populations contributing to the
flux.

Once de-reddened, the rest-frame (B-J) colours reveal young stellar
populations with ages from 0 to 2 Gyrs, significantly bluer than
those of early-type galaxies, indicative of on-going SF.
The (V-K) de-reddened colours show a dependence on redshift:
while at $z<0.6$ they appear blue, those for galaxies at
$z>0.7$ are constant and quite red on average ($2<V-K<3$),
and as red as those of the early-type population investigated
by FA98.

We warn that translation to age-distributions is subject to the 
uncertainties in the evaluation of the effective
extinction (see also next Section).
However, the similarity in the intrinsic V-K colours of galaxies 
independent of morphology does indeed support a common age distribution
for the spheroidal stellar components in Elliptical/S0s and in
spiral bulges, something predicted by the hierarchical formation 
scenario mentioned in Section 1.

\begin{figure}
\begin{center}
 \resizebox{9cm}{!}{\includegraphics{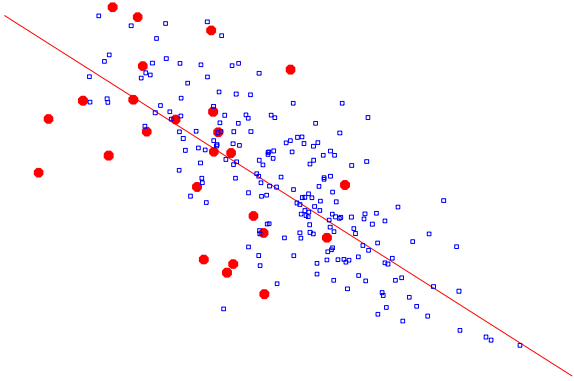}}
\end{center}
\caption{Kormendy relation in the B band, i.e., the
average surface brightness versus effective radius 
for the {\it bona-fide} spirals in our high-z sample
(filled cirles), compared with the data from a local spiral sample 
based on the RC3 (open squares).  $<\mu_B>$ was obtained for 
our distant sources applying the K-correction and
a correction for the internal absorption. }
\label{kormendy}
\end{figure}

Finally, we report in Figure \ref{kormendy} our measured average surface
brightness in the B band ($<\mu_B>$) versus effective radius $R_e$ for the {\it bona-fide}
spirals in our high-z sample, compared with data from a local sample 
based on the RC3. $<\mu_B>$ was computed for our distant sources applying
only the K-correction and the cosmological scaling factors.
A further correction has been applied taking into account
the effects of internal absorption.
While the largest values of $R_e$ shown by local galaxies
are missed by our high-z sample because we are not sampling the rare
population of large size galaxies in the HDF limited space volume,
there is no evidence of a significant offset in $<\mu_B>$ between the
local and distant spirals within the large observed spreads in the
data. In particular, the lowest surface brightness galaxies ($<\mu_B> \simeq
22-23$) in our sample are highly inclined, probably extinguished, 
objects. Note that the same relation fitting the Kormendy relation
for E/S0 galaxies also fits data for spirals.

\subsection{A tentative physical characterization of late-type galaxies 
in the HDF-N}

Though aware of the uncertainties inherent in the spectral modelling
of gas-rich systems due to the uncertain extinction, 
neverthless we attempt here to estimate some basic physical
parameters of these sources, or at least to provide some boundary values 
as found by application of our vast model grid. 

We report in Table 1 the formal best-fit solutions obtained from fitting the observed 
SEDs of our sample galaxies:
$A_V$: V band effective extinction;
$M_{tot}$: total baryonic mass divided by $10^{11}$ solar masses;
$SFR$: observed SFR in solar masses per year. 

\begin{figure}
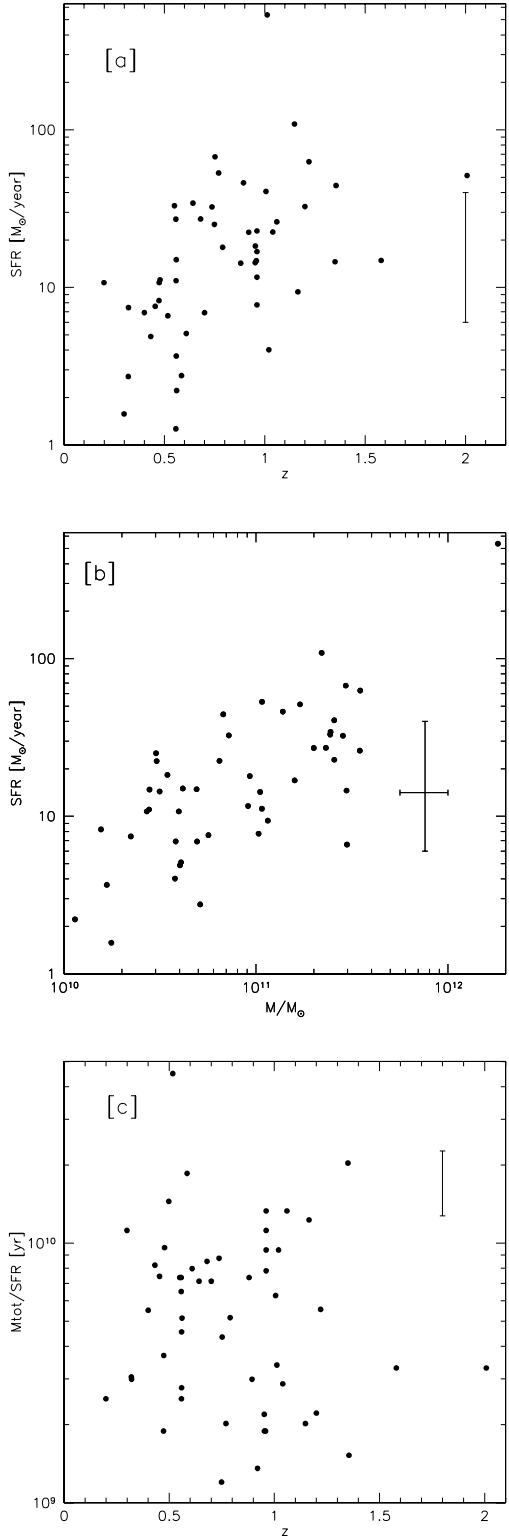

 \resizebox{7cm}{!}{\includegraphics{9985.f13}}
 \resizebox{7cm}{!}{\includegraphics{9985.f14}} 
 \hfill 
 \resizebox{7cm}{!}{\includegraphics{9985.f15}} 
\caption{Panel $a$: Distribution of the on-going star formation rate SFR versus
redshift, for the best-fit solutions.  
Panel $b$: SFR against the total baryonic mass for best fit solutions.
The mean uncertainty on the mass is reported. Panel $c$: Ratio of the total 
baryonic mass normalized to the on-going 
rate of SF for each objects (based on best-fit solutions).
This ratio gives an estimate of the timescale for the conversion of gas into
stars, showing a substantial spread from a few to 20 Gyrs.
The error bars correspond to the mean 90\% uncertainty, providing the range 
of variation for two extreme solutions (the youngest and the oldest one with
$\chi^2=\chi^2_{best}+5$).}
\label{zSFRbf}
\end{figure}

Figure\ \ref{zSFRbf}[a] plots the rate of ongoing star-formation SFR based on
best-fit solutions versus redshift for our sample galaxies.
The values derived in our analysis have a median around $SFR=10\ M_{\odot}/yr$. 
Only one peculiar object (source number 2 in Table 1) 
shows an extreme value of SF 
(above $\sim 500 M_{\odot}/yr$). 
It is an apparently
normal giant spiral viewed face-on, for which our spectral fit predicts
a large extinction $A_V=2.2$. A more standard extinction value ($A_V=1.3$), 
still providing an acceptable fit, 
would still correspond to a large value of SFR$\sim 200 M_{\odot}/yr$.

The apparent scaling of SFR with z in Fig.\ref{zSFRbf}[a] may be explained as 
mostly a selection effect concerning the luminousity of our objects. 
On the other hand, Figure\ \ref{zSFRbf}[b]
indicates that the star formation rate is on average proportional to the
intrinsic baryonic mass, such that galaxies with
higher SFR are typically those more massive. 
By looking at higher redshifts means to observe
only the more luminous sources, those with larger masses.
Our K-band selection then operates largely on the stellar mass.

Any dependence on redshift disappears when we normalize SFR to the
baryonic mass, as it is done in Figure\ \ref{zSFRbf}[c].
The ratio $M_{tot}/SFR$ appearing in Fig.\ref{zSFRbf}[c]
gives the timescale for the formation of stars in 
our late--type galaxy sample.
The latter does not reveal characteristics of violent starburst, if we
consider our observed timescales required to convert all gas in stars:
these range from 1 up
to 20 Gyrs, and indicate a moderate star formation activity 
for the present K-selected field galaxies.

\subsection{Constraints on the global star formation history: contributions
of late-type and early-type field galaxies}

A most popular way to represent the evolutionary properties of a
population of cosmic sources is through the plot of the total luminosity
density (or the stellar formation and metal production rates) 
in the comoving volume (Madau et al. \cite{madau}; Lilly et al. 
\cite{lillyb}).
When referred to the average galaxy
population in the field, this function was shown to drastically increase from
the present time back to redshift $z\sim 1$, and to flatten off above.

The separate contribution of galaxies with early-type morphologies
to the global star-formation rate per comoving volume $\Psi(z)$
[$M_{\odot}~yr^{-1}~Mpc^{-3}$]
has been estimated by FA98 using the complementary
sample in the HDFN and population synthesis results. The outcome
was that early-types contribute significantly to the total 
$\Psi(z)$ mostly at $z>1$, their fractional contribution
decreasing very fast at lower $z$.

A first reason to perform a similar computation on the complementary
sample of spirals and irregulars is to compare the two histories
of SF. A further reason of interest to have the full complete sample 
processed comes from recent reports claiming evidence for a more
gradual decline of the galaxy ultraviolet luminosity density
at $z<1$ (Cowie et al. \cite{cowie}; Treyer et al.
\cite{treyer}), taken as an indication of a modest evolution of the 
rate of SF during the last $\sim 10$ Gyrs of the galaxy cosmic history.

An independent assessment, accounting 
for dust extinction and exploiting the observed baryonic
mass function in stars through a full spectro-photometric fit
to the SED's, would then be clearly welcome.
We remember that, whereas this computation is relatively 
straightforward for the classified ellipticals/S0 due to the
lack of an ISM complicating the stellar population-synthesis fit, modelling 
gas rich late-types presents more severe problems due to the presence of dust.
We will see later, however, that the corresponding uncertainties
tend to average out in the integrated form of the $\Psi(z)$ function,
providing a relatively robust result.

We defer to the paper by Franceschini et al.\ (1998) for all
details of the computation. To remind here only the basic steps,
for all 52 objects in our complete sample we computed,
within our grids of synthetic spectra, the younger and more 
extinguished solution. In the same way we determined
the older solution less affected by absorption. 
We computed the available comoving volumes $V_{max}$ within which the
object would still be visible above the sample flux limit (Lilly et
al. \cite{lilly}, \cite{lillyb}). 
The contribution of each galaxy to the global SFR has been estimated 
by dividing the time dependent SF rate (derived from the two fits) 
by $V_{max}$. A correction to the comoving SF rate is then applied
for the portion of the luminosity function not sampled by the present
survey.  Such correction is based on the $K$-band
luminosity function discussed by Connolly et al. (\cite{connolly}).
The global SFR density $\Psi(t)$,
is the summed contribution by all galaxies in our sample.

The result appears in Figure\ \ref{sfrtot} in the form of the comoving
rate of star formation $\Psi(z)$ 
versus redshift for the sample considered here
(dot-dash line), compared with the evolutionary path for early-type galaxies
(dotted lines). The results in panel (a) and (b) correspond to
the two extreme acceptable (at 90\% of confidence) 
spectral solutions for each object, the one most extinguished and younger
for panel (a), and the older less extinguished for panel (b).

\begin{figure}
\begin{center}
 \resizebox{9cm}{!}{\includegraphics{9985.f16}}
\caption{Comoving volume star-formation rate density $\Psi(t)$ 
as a function of redshift for field galaxies.
The contribution of late-types to the cosmic SFR (dot dashed line) derived
from our sample is compared in the two cases with the evolutionary path for
early-type galaxies studied by FA98 (dotted line). The solid 
line corresponds to the total amount of SF density in the field. 
 The panels correspond to two different extreme solutions
(see text for details): the younger and more extinguished (panel [a]), the 
older less affected by dust absorption (panel [b]). 
The data reported are from Lilly et al. (\protect\cite{lillyb}) and
Connolly et al. (\protect\cite{connolly}).}
\label{sfrtot}
\end{center}
\end{figure}

\begin{figure}
\begin{center}
 \resizebox{9cm}{!}{\includegraphics{9985.f18}}
\end{center}
\caption{The total star formation rate density per unit comoving volume
for field galaxies (late+early-types) is compared
with the prediction of Cowie et al. (\protect\cite{cowie}), who found a dependence of
$\Psi(z)$ on $z$ in the range $0.2<z<1.$ of the form $\Psi(z) \propto
(1+z)^{1.5}$ (continous line).
The shaded region is bracketed by the two solutions based on the
younger-high extinction (upper histogram) and the older less
extinguished (lower histogram) models. See also caption to Fig.\ \ref{sfrtot}.}
\label{starform}
\end{figure}

The disk--dominated and the irregular galaxies in the present sample
display an evolutionary behaviour different from that of 
bulge--dominated objects.
The former appear to form actively stars well below $z=1$, whereas the rate
of SF for the latter is high at $z>1$ but converges very fast at lower $z$.
Our result for the disk and irregular galaxies is quite consistent with
those by Brinchmann et al. (\cite{brinchmann}, their fig. 15), 
in showing a comoving SFR
modestly increasing between $z=0.3$ and 1. On the contrary, our results
differ significantly from Brinchmann et al. as far as the early-type 
systems are considered (in their case E/S0 have a flat $\Psi(t)$ in the
same z-interval): we explain this as due to the very different procedures
adopted to measure the function $\Psi(z)$, in our case it was a global
fit to the UV-optical-NIR SED, in their case the use of the OII EW as a
tracer of SF. Indeed,
the latter should more likely trace a negligible residual of SF due
to low-level merging activity or stellar recycling, than the global
history of SF in these galaxies.

Note that, despite the large uncertainties on the single
object, the overall result is fairly well constrained between the two
extreme solutions depicted by the shaded region in Figure \ref{starform}. 
This is due to a sort of compensation intervening in the adopted solutions: 
the younger-more extinguished one
tends to have a more intense ongoing SF activity but less protracted in time,
while the contrary happens for the older less-extinguished solutions. 
In other words, the baryonic mass already converted into stars and sampled
by the near-IR (JHK) flux measurements as a function of redshift, provides
a more robust evaluation of the evolutionary SFR than the 
instantaneous SFR mapped by the short-wavelength flux. 
In a sense, the errorbar appearing in Fig.\ \ref{zSFRbf} does not translates
into a similarly large uncertainty in the prediction of Fig.\ \ref{starform},
because the ongoing rate of SF (SFR in panel [a] of Fig.\ \ref{zSFRbf}) and the timescale
of SF ($M_{tot}/SFR$ in panel [c]) scale inversely to 
the galaxy mass function observed at various redshifts.

This prompted us to compare our results with those published
by Cowie et al. (\cite{cowie}). This is done in Figure\ \ref{starform} where our
results appear as the shaded region, which is bracketed by the two
solutions based on the younger-high extinction (upper histogram) and
the older less extinguished (lower histogram) models.
The continuous line is a polynomial function [$\Psi(z) \propto
(1+z)^{1.4}$] quoted by Cowie et al. (\cite{cowie}) as best-fitting their
and Treyer's et al. (\cite{treyer}) data on the time-dependent
UV luminosity density. Within the uncertainties, our results are in
quite better consistency with the Cowie et al. (1999) evolutionary
law than with the dataset compiled by Madau et al. (\cite{madau}), based
on the CFRS (Lilly et al. \cite{lillyb}) and the low-z H$_{\alpha}$ survey by
Gallego et al. (\cite{gallego}).

While some discussions can be found in Cowie et al. (\cite {cowie}) about possible
origins for this discrepancy and on the consequences on this
new evaluation of the evolutionary SFR, we only
take note here of the nice agreement between our results and those
of Cowie et al. (\cite{cowie}), based on quite
indipendent grounds.

It is remarkable that UV and near IR selected galaxy samples show
such similar evolution of the comoving SFR density $\Psi(z)$. 

\section{DISCUSSION AND CONCLUSIONS}

With the main goal to investigate systematic differences between 
early-type and late-type galaxies -- as for colours, 
redshift distributions, and ages of the dominant stellar populations --
we have analyzed a morphologically-selected complete sample of 52 
spiral and irregular
galaxies in the Hubble Deep Field North with total K-magnitudes
brighter than K=20.47 and typical redshifts from $z\sim 0.5$ to 1.5. 
The sample makes use of total photometry in the UBVI bands 
from HST and the JHK bands from ground, all carefully tested 
with an extensive set of Monte Carlo simulations.

The present sample exploits in particular the
ultimate imaging quality achieved by HST in this field, allowing
us to disentangle among galaxy morphologies,
based on accurate profiles of the surface brightness distributions.

Our analysis makes also use of an exhaustive set of modellistic spectra
accounting for a variety of physical and geometrical situations
for the stellar populations, the dusty ISM, and relative assemblies.
The high photometric quality and wide spectral coverage
allowed us to estimate accurate photometric
redshifts for 16 objects lacking a spectroscopic measurement.

We have also carefully evaluated all plausible systematic effects 
of the selection, in particular the redshift cutoff implied by
the limiting surface-brightness achievable in the reference K band image. 

A warning is in order, in any case, about the general conclusions derived from 
our sample of K-selected galaxies: they should be treated with caution, due to
the very small field of view and modest spatial sampling of the present survey.
Ferguson et al. (2000) and Eisenhardt et al. (2000) estimate that the number of 
$L^*$ galaxies in the total HDF co-moving volume between z=1 and z=2 is 
only a few dozens.
Considering also the strong clustering inferred for Lyman break
galaxies (Adelberger et al. 1998), statistical fluctuations imply
large uncertainties on any conclusions based on samples like the HDF,
untill more substantial surveys to similar depths will be made available. 

Three the main results of our study.

\begin{itemize}

\item
The sample galaxies are distributed in redshift up to $z=1.4$,
but appear to be significantly missing above, compared
with evolutionary models assuming standard recipes for the 
luminosity evolution and a substantial redshift of formation.
We reported a similar finding in our previous study 
of early-type galaxies in the same area (Franceschini et al. 1998).
Our conclusion is that, either the area has some peculiarities,
or the underlying mass function for galaxies of all morphological
kinds has a global decline at these high redshifts.
Confirmation of this result will require more substantial
sky areas to be surveyed to similar depths by large telescopes.

\item
Differences between early- and late-types are apparent
in the rest-frame colour distributions and the evolutionary 
star-formation rates per unit volume.
In particular, the short-wavelength rest-frame colours B-J,
once dust reddening is taken into account, appear quite
significantly bluer for late- than for early- galaxy types.
On the contrary, the longer-wavelength colors V-K appear to
be very similar for the two morphological classes.
We interprete this as an indication that, while the stellar
mix in spirals/irregulars includes young newly formed 
populations, less apparent in E/S0, 
the underlying older age component traced by the V-K colors
has quite a similar origin and age distribution for the 
two galaxy categories.

We warn, however, that the complication 
in spectro-photometric modelling introduced by dust-extinction 
in the gas-rich systems prevents to reach conclusive results
on the source by source basis. Only future long-wavelength 
IR observations, from space (SIRTF, FIRST, NGST) and
from ground (10 m class telescopes in the mid-IR and interferometers
in the sub-mm), will allow to break down the age/extinction degeneracy.

\item
We found that an integrated quantity like the 
comoving-volume star-formation rate density as a function of redshift  
$\Psi(z)$ is much less affected by the uncertainties related to the
dust distribution. The reason for this is mostly in the fact
that our analysis is strongly constrained by the evolutionary
baryonic mass function in stars traced by the near-IR galaxy luminosities,
the estimate of the baryon mass at any redshifts being much more 
robust than that of the instantaneous rate of star formation
(see Sect. 4.5).

By combining this with
the early-type galaxy sample previously studied by FA98, we find 
a shallower dependence of $\Psi(z)$
on $z$ between $z=0.2$ and $z=1.5$ than found by Lilly et al. (1996),
i.e. $\Psi(z)\propto (1+z)^{1.4}$ rathen than 
$\Psi(z)\propto (1+z)^{4}$ as in Lilly et al. (\cite{lillyb}).
In this redshift interval our observed  $\Psi(z)$ turns out to
roughly agree with results published by Cowie et al. (\cite{cowie})
and Treyer et al. (\cite{treyer}).

Our present results, based for the first time on a careful modelling 
of the whole UV-optical-NIR Spectral Energy Distributions of galaxies,
then support a revision of the Lilly-Madau plot at low-redshifts for 
UV- and K-band selected galaxies.
UV-selected and near infrared selected galaxy samples display a remarkably
similar evolution of the comoving SFR density $\Psi(z)$, at $z<1$.

\end{itemize}

The three above findings seem to
favour the general scheme of hierarchical assembly for
the formation of bright galaxies, envisaging their progressive build up
during a substantial fraction of the Hubble time. After all, this
is the most physically motivated present description. 

In this context, a warning is in order concerning some published 
specializations of the Cold Dark Matter cosmogonic scheme, 
predicting that spheroidal galaxies in the field form at low
redshift ($z<1$) from merging of spirals (Kauffmann \& Charlot 
\cite{kauffmannb}).
This prediction is not supported by our results in Fig.\ \ref{sfrtot},
where Ellipticals and S0s appear to have been mostly formed
at $z=1$, whereas spirals/irregulars keep some sustained SF activity at
$z<1$. This result suggests that merging of spirals to form ellipticals
at low redshifts cannot be a dominant process, quite in
agreement with what found by Brinchmann \& Ellis (\cite{brinchmannb}).

It is conceivable, however, that minor modifications of the CDM
hierachical scheme (e.g. in terms of different assumptions about the 
cosmological parameters $\Omega_m$, $\Omega_{\Lambda}$) 
can explain the observed differences between
morphological types. In our view, these differences in the population 
histories are quite probably related to the presence of different environments  
at different
densities (and consequently different cosmic timescales of
formation) in what we call the "field". 
In particular, moderately high-density environments
(typically galaxy groups, as found very numerous in the spectroscopic
survey by Cohen et al. \cite{cohenb}), with an accelerated cosmic 
timescale of evolution and fast gas consumption, 
mix with truly low-density environments,
where the transformation of primordial gas into stars slowly
progresses during the whole Hubble time. 
We believe that the two galaxy morphologies analyzed 
in the present paper and in FA98 trace such different environments
in the universe.

\begin{table*}
\caption{Photometric data on the sample galaxies}
\begin{tabular}{|c|c|c|c|c|c|c|c|c|c|c|c|c|c|c|c|}
\hline
id&$\alpha$&$\delta$&~&$r_e(")$&$U_{AB}$&$B_{AB}$&$V_{AB}$&$I_{AB}$&$J_{corr}$&$H_{corr}$&$K_{corr}$&$z$& $A_V$ & $M_{tot}$ & SFR\\
\hline
~&$s$&$'$&$''$&~&~&~&~&~&~&~&~&~&~&~&\\
\hline                                                          
0   &   56.65   &   12  &   45.60   &   1.21    &   24.91   &   22.49   &   21.09   &   20.05   &   18.26   &   17.48   &   16.79   &   0.517   &   0.17    &   2.97    &   6.61    \\
1   &   51.08   &   13  &   20.73   &   1.12    &   21.69   &   20.52   &   19.95   &   19.63   &   18.40   &   17.87   &   17.30   &   0.199   &   1.03    &   0.26    &   10.73   \\
2   &   46.15   &   11  &   42.05   &   0.34    &   23.40   &   22.90   &   22.40   &   20.80   &   19.17   &   18.38   &   17.37   &   1.012   &   2.25    &   18.19   &   536.4   \\
3   &   53.90   &   12  &   54.05   &   0.60    &   23.74   &   22.80   &   21.89   &   20.88   &   19.12   &   18.30   &   17.47   &   0.642   &   1.16    &   2.44    &   34.28   \\
4   &   43.96   &   12  &   50.13   &   0.31    &   23.90   &   22.87   &   21.83   &   21.01   &   19.27   &   18.46   &   17.59   &   (0.55)  &   1.74    &   2.43    &   33.01   \\
5   &   44.58   &   13  &   4.66    &   0.54    &   25.24   &   23.63   &   22.23   &   21.24   &   19.29   &   18.40   &   17.67   &   (0.68)  &   1.15    &   2.31    &   27.18   \\
6   &   51.78   &   13  &   53.73   &   0.49    &   23.92   &   22.97   &   21.97   &   21.09   &   19.42   &   18.51   &   17.70   &   0.557   &   1.56    &   1.99    &   27.12   \\
7   &   42.91   &   12  &   16.26   &   0.52    &   22.86   &   22.22   &   21.32   &   20.74   &   19.20   &   18.53   &   17.90   &   0.454   &   0.28    &   0.56    &   7.58    \\
8   &   49.75   &   13  &   13.09   &   0.63    &   24.99   &   23.54   &   22.29   &   21.49   &   19.07   &   18.56   &   18.00   &   0.478   &   1.34    &   1.07    &   11.17   \\
9   &   50.25   &   12  &   39.72   &   0.49    &   22.59   &   22.08   &   21.25   &   20.69   &   19.3    &   18.67   &   18.00   &   0.474   &   0.37    &   0.39    &   10.73   \\
10  &   41.95   &   12  &   5.41    &   0.47    &   23.34   &   22.60   &   21.67   &   21.03   &   19.42   &   18.78   &   18.04   &   0.432   &   0.34    &   0.40    &   4.88    \\
11  &   45.85   &   13  &   25.81   &   0.85    &   23.23   &   22.30   &   21.45   &   20.95   &   19.45   &   18.86   &   18.13   &   (0.4)   &   0.48    &   0.38    &   6.92    \\
12  &   43.18   &   11  &   48.05   &   0.47    &   26.05   &   24.95   &   23.91   &   22.45   &   20.14   &   19.29   &   18.23   &   (1.06)  &   0.89    &   3.47    &   26.09   \\
13  &   51.72   &   12  &   20.18   &   0.34    &   24.85   &   23.34   &   22.24   &   21.55   &   19.87   &   19.12   &   18.31   &   0.299   &   0.95    &   0.17    &   1.57    \\
14  &   42.72   &   13  &   7.26    &   0.33    &   25.61   &   24.58   &   23.22   &   22.19   &   20.00   &   19.21   &   18.32   &   (0.737) &   1.73    &   2.83    &   32.42   \\
15  &   47.04   &   12  &   34.96   &   0.40    &   22.84   &   22.19   &   21.46   &   21.03   &   19.66   &   19.05   &   18.34   &   0.321   &   1.00    &   0.22    &   7.45    \\
16  &   49.51   &   14  &   6.77    &   0.34    &   24.17   &   23.58   &   22.82   &   21.83   &   20.17   &   19.23   &   18.41   &   0.752   &   2.07    &   2.93    &   67.43   \\
17  &   49.51   &   12  &   20.11   &   0.61    &   25.58   &   25.16   &   24.04   &   22.46   &   20.34   &   19.40   &   18.53   &   0.961   &   1.03    &   2.55    &   22.83   \\
18  &   41.42   &   11  &   42.89   &   0.33    &   26.18   &   25.13   &   24.81   &   24.16   &   20.69   &   19.66   &   18.60   &   1.320   &   1.03    &   1.58    &   16.88   \\
19  &   58.76   &   12  &   52.35   &   0.85    &   23.18   &   22.52   &   21.72   &   21.31   &   19.92   &   19.34   &   18.77   &   0.320   &   0.38    &   0.08    &   2.72    \\
20  &   55.58   &   12  &   45.43   &   0.47    &   24.18   &   23.63   &   22.97   &   21.97   &   20.35   &   19.65   &   18.86   &   0.790   &   0.70    &   0.92    &   17.98   \\
21  &   50.47   &   13  &   16.16   &   0.64    &   25.46   &   24.62   &   23.74   &   22.60   &   20.68   &   19.95   &   18.90   &   (0.88)  &   1.65    &   1.05    &   14.25   \\
22  &   57.33   &   12  &   59.63   &   0.44    &   22.83   &   22.53   &   21.82   &   21.38   &   20.10   &   19.42   &   18.93   &   0.473   &   0.47    &   0.15    &   8.25    \\
23  &   41.31   &   11  &   40.87   &   0.33    &   27.02   &   24.70   &   23.61   &   22.47   &   20.57   &   19.73   &   19.01   &   0.585   &   0.67    &   0.51    &   2.75    \\
24  &   38.44   &   12  &   31.35   &   0.59    &   26.76   &   25.15   &   23.74   &   22.67   &
    20.85   &       20.03   &       19.17   &      (0.7)    &       1.06    &       0.49    &       6.90    \\  
25  &   49.24   &   11  &   48.38   &   0.32    &   25.46   &   24.93   &   24.16   &   22.96   &   20.67   &   19.98   &   19.18   &   0.961   &   0.23    &   1.03    &   7.75    \\
26  &   48.62   &   12  &   15.81   &   0.47    &   26.70   &   26.20   &   25.20   &   23.60   &   21.23   &   20.23   &   19.21   &   (1.35)  &   0.96    &   2.95    &   14.55   \\
27  &   49.45   &   13  &   16.58   &   0.28    &   25.41   &   24.44   &   24.02   &   23.27   &   21.18   &   20.22   &   19.22   &   (1.006) &   1.74    &   2.55    &   40.67   \\
28  &   48.12   &   12  &   14.90   &   0.88    &   24.26   &   23.83   &   23.54   &   22.64   &   20.53   &   20.02   &   19.36   &   0.961   &   0.27    &   0.90    &   11.62   \\
29  &   54.10   &   13  &   54.35   &   0.45    &   24.35   &   23.84   &   23.37   &   22.43   &   21.01   &   20.17   &   19.41   &   0.894   &   1.68    &   1.37    &   46.13   \\
30  &   52.70   &   13  &   55.49   &   0.27    &   23.96   &   23.22   &   23.06   &   22.71   &   20.64   &   20.08   &   19.41   &   1.355   &   0.54    &   0.67    &   44.33   \\
31  &   38.99   &   12  &   19.63   &   0.45    &   24.14   &   23.67   &   22.95   &   22.22   &   20.81   &   20.17   &   19.57   &   (0.77)  &   1.51    &   1.07    &   53.23   \\
32  &   44.19   &   12  &   47.90   &   0.39    &   22.99   &   22.65   &   22.13   &   21.65   &   20.50   &   19.91   &   19.58   &   .558   &    0.50    &   0.27    &   11.04   \\
33  &   49.01   &   12  &   20.86   &   0.18    &   23.87   &   23.54   &   23.23   &   22.44   &   20.89   &   20.53   &   19.68   &   0.953   &   0.47    &   0.34    &   18.31   \\
34  &   39.56   &   12  &   13.83   &   0.44    &   28.20   &   26.49   &   25.15   &   23.79   &   21.62   &   20.53   &   19.69   &   (1.22)  &   2.07    &   3.48    &   62.76   \\
35  &   48.27   &   13  &   13.8    &   0.29    &   26.53   &   25.90   &   25.30   &   24.07   &   21.51   &   20.36   &   19.75   &   (1.165) &   0.57    &   1.15    &   9.37    \\
36  &   53.45   &   12  &   34.52   &   0.34    &   24.97   &   24.25   &   23.50   &   22.81   &   21.47   &   20.83   &   19.80   &   0.559   &   2.08    &   0.41    &   15.00   \\
37  &   55.53   &   13  &   53.48   &   0.70    &   24.04   &   23.49   &   23.23   &   22.65   &   21.27   &   20.48   &   19.80   &   1.148   &   1.70    &   2.19    &   108.8   \\
38  &   49.58   &   14  &   14.63   &   0.48    &   25.40   &   24.00   &   23.80   &   22.70   &   21.37   &   20.75   &   19.94   &   (0.92)  &   0.82    &   0.30    &   22.39   \\
39  &   57.67   &   13  &   15.32   &   0.39    &   24.43   &   23.94   &   23.57   &   22.82   &   21.29   &   20.71   &   19.97   &   0.952   &   0.46    &   0.31    &   14.37   \\
40  &   55.50   &   14  &   2.71    &   0.55    &   26.06   &   24.87   &   23.87   &   22.99   &   21.48   &   20.75   &   20.01   &   0.559   &   1.07    &   0.16    &   3.66    \\
41  &   48.79   &   13  &   18.35   &   0.68    &   24.27   &   23.93   &   23.50   &   22.76   &   21.32   &   20.27   &   19.66   &   0.749   &   1.85    &   0.30    &   25.12   \\
42  &   47.19   &   14  &   14.18   &   0.22    &   26.70   &   25.74   &   24.59   &   23.53   &   21.84   &   20.63   &   20.02   &   0.609   &   1.83    &   0.40    &   5.10    \\
43  &   57.21   &   12  &   25.83   &   0.51    &   24.17   &   24.02   &   23.91   &   22.45   &   21.34   &   20.60   &   20.05   &   0.561   &   0.32    &   0.11    &   2.21    \\
44  &   48.34   &   14  &   16.63   &   0.16    &   25.86   &   23.97   &   23.73   &   23.43   &   21.64   &   20.53   &   20.11   &   2.008   &   0.33    &   1.69    &   51.3    \\
45  &   47.78   &   12  &   32.93   &   0.28    &   25.21   &   24.73   &   24.27   &   23.34   &   21.47   &   21.12   &   20.12   &   (1.04)  &   0.98    &   0.64    &   22.47   \\
46  &   52.87   &   14  &   5.11    &   0.29    &   25.70   &   24.84   &   24.06   &   23.25   &   21.38   &   20.89   &   20.21   &   0.498   &   0.16    &   0.08    &   0.57    \\
47  &   52.02   &   14  &   0.91    &   0.33    &   25.25   &   24.55   &   23.72   &   23.02   &   21.48   &   20.74   &   20.22   &   0.557   &   0.23    &   0.08    &   1.26    \\
48  &   48.58   &   13  &   28.35   &   0.47    &   25.33   &   24.30   &   23.95   &   23.00   &   21.22   &   21.08   &   20.24   &   0.958   &   0.68    &   0.27    &   14.77   \\
49  &   44.64   &   12  &   27.39   &   0.24    &   25.79   &   24.28   &   24.11   &   23.73   &   21.63   &   20.89   &   20.29   &   (1.58)  &   0.18    &   0.49    &   14.84   \\
50  &   44.45   &   11  &   41.82   &   0.62    &   24.43   &   24.46   &   24.52   &   23.88   &   21.05   &   21.25   &   20.42   &   1.020   &   0.17    &   0.37    &   4.02    \\
51  &   56.13   &   13  &   29.74   &   0.39    &   24.38   &   24.13   &   23.97   &   23.34   &   21.38   &   21.30   &   20.45   &   (1.2)   &   0.80    &   0.72    &   32.59   \\
41  &   48.79   &   13  &   18.35   &   0.68    &   24.27   &   23.93   &   23.50   &   22.76   &   21.32   &   20.27   &   19.66   &   0.749   &   1.85    &   0.30    &   25.12   \\
42  &   47.19   &   14  &   14.18   &   0.22    &   26.70   &   25.74   &   24.59   &   23.53   &   21.84   &   20.63   &   20.02   &   0.609   &   1.83    &   0.40    &   5.10    \\
43  &   57.21   &   12  &   25.83   &   0.51    &   24.17   &   24.02   &   23.91   &   22.45   &   21.34   &   20.60   &   20.05   &   0.561   &   0.32    &   0.11    &   2.21    \\
44  &   48.34   &   14  &   16.63   &   0.16    &   25.86   &   23.97   &   23.73   &   23.43   &   21.64   &   20.53   &   20.11   &   2.008   &   0.33    &   1.69    &   51.3    \\
45  &   47.78   &   12  &   32.93   &   0.28    &   25.21   &   24.73   &   24.27   &   23.34   &   21.47   &   21.12   &   20.12   &   (1.04)  &   0.98    &   0.64    &   22.47   \\
46  &   52.87   &   14  &   5.11    &   0.29    &   25.70   &   24.84   &   24.06   &   23.25   &   21.38   &   20.89   &   20.21   &   0.498   &   0.16    &   0.08    &   0.57    \\
47  &   52.02   &   14  &   0.91    &   0.33    &   25.25   &   24.55   &   23.72   &   23.02   &   21.48   &   20.74   &   20.22   &   0.557   &   0.23    &   0.08    &   1.26    \\
48  &   48.58   &   13  &   28.35   &   0.47    &   25.33   &   24.30   &   23.95   &   23.00   &   21.22   &   21.08   &   20.24   &   0.958   &   0.68    &   0.27    &   14.77   \\
49  &   44.64   &   12  &   27.39   &   0.24    &   25.79   &   24.28   &   24.11   &   23.73   &   21.63   &   20.89   &   20.29   &   (1.58)  &   0.18    &   0.49    &   14.84   \\
50  &   44.45   &   11  &   41.82   &   0.62    &   24.43   &   24.46   &   24.52   &   23.88   &   21.05   &   21.25   &   20.42   &   1.020   &   0.17    &   0.37    &   4.02    \\
51  &   56.13   &   13  &   29.74   &   0.39    &   24.38   &   24.13   &   23.97   &   23.34   &   21.38   &   21.30   &   20.45   &   (1.2)   &   0.80    &   0.72    &   32.59   \\
\hline
\end{tabular}
\end{table*}


\begin{thebibliography}{}

\bibitem[1997]{adelberger}
  Adelberger, K.L., Steidel, C.C., Giavalisco, M., Dickinson, M.E., Pettini, M.,
  Kelog, M. 1998, ApJ 505, 18

\bibitem[1996]{bertin}
 Bertin, E., Arnouts, S., 1996, A\&AS 117, 393

\bibitem[1998]{brinchmann}
 Brinchmann, J., Abraham, D.S., Tresse, L., Ellis, R.S., Lilly, S. et al., 1998, ApJ 499, 112

\bibitem[1998]{brinchmannb}
 Brinchmann, J. and Ellis, 2000, astro-ph/0005120

\bibitem[1996]{cohen}
 Cohen, J.G., Cowie, L.L., Hogg, D.W., Songaila, A., Blandford, R.,
 Hu, E.M., Shopbell, P., 1996, AJ 471, L5

\bibitem[1999]{cohenb}
 Cohen, J.G., Blandford, R., Hogg, D.W., Pahre, M.A., Shopbell, P.L., 1999, ApJ 512, 30

\bibitem[1997]{connolly}
 Connolly, A.J., Szalay, A.S., Dickinson, M., Subbarao, M.U., Brunner, R.J., 1997, ApJ 486, L11

\bibitem[1999]{cowie}
 Cowie, L.L., Songaila, A., Barger, A.J., 1999, AJ 118, 603

\bibitem[1996]{cowieb}
 Cowie, L.L. et al., 1996, http://www.ifa.hawaii.edu/cowie/hdf.html

\bibitem[1997]{dickinson}
 Dickinson, M. et al., 1997, http://archive.stsci.edu/hdf/hdfirim.html

\bibitem[2000]{eisenhardt}
 Eisenhardt, P., Elston, R., Stanford, S.A., Dickinson, M., Spinrad, H., Stern, D., Dey. A.,
 2000, astro-ph/0002468

\bibitem[1997]{ellis}
 Ellis, R.S., 1997, ARA\&A 35, 389


\bibitem[1998]{soto}
 Fernandez-Soto, A., Lanzetta, K.M., Yahil, A., 1998, http://bat.phys.unsw.edu.au/fsoto/hdf

\bibitem[1998]{franceschini}
 Franceschini, A., Silva, L., Fasano, G., Granato, G.L., Bressan, A., Arnouts, S.,
Danese, L., 1998, ApJ 506, 600

\bibitem[1997]{gardner}
 Gardner, J.P., Sharples, R.M., Frenk, Carrasco, B.E., 1997, ApJ 480, L99


\bibitem[1995]{gallego}
Gallego, J., Zamorano, J., Aragon-Salamanca, A., Rego, M., 1995, ApJ 455, L1

\bibitem[1998]{kauffmannb}
  Kauffmann, G., Charlot, S., 1998, MNRAS 297, L23

\bibitem[1995]{lilly}
 Lilly, S.J., Tresse, L., Hammer, F., Crampton, D., Le Fevre, O., 1995, ApJ 455, 108

\bibitem[1996]{lillyb}
 Lilly, S.J., Le Fevre, O., Hammer, F., Crampton, D., 1996, ApJ 460, L1

\bibitem[1996]{madau}
 Madau, P., Ferguson, H.C., Dickinson, M.E., Giavalisco, M., Steidel, C.C., Fruchter, A.,
 1996, MNRAS 283, 1388

\bibitem[1983]{oke}
 Oke, J.B., Gunn, J.E., 1983, ApJ 266, 713

\bibitem[2000]{rodighiero}
 Rodighiero, G., Franceschini, A. and Fasano, G., 2000, in preparation

\bibitem[1998]{silva}
 Silva, L., Granato, G.L., Bressan, A., Danese, L., 1998, ApJ 509, 103

\bibitem[1998]{treyer}
 Treyer, M.A., Ellis, R.S., Milliard, B., Donas, J., Bridges, T.J., 1998, MNRAS 300, 303

\bibitem[1996]{williams}
 Williams, R.E., Blacker, B., Dickinson, M., Dixon, W.V., Ferguson, H.C.
 et al., 1996, AJ 112, 1335

\bibitem[1999]{wolfe}
 Wolfe, A.M., 1999, American Astronomical Society, Meeting 194, \#63.07


\end{thebibliography}
\end{document}